\documentclass{elsart}
\usepackage{graphics}
\usepackage{graphicx}
\usepackage{epsfig}
\usepackage{amssymb}
\begin{document}

\begin{frontmatter}

\title{The $\mathbf{pp{\to}K^+n\Sigma^+}$ reaction near threshold}

\author[PNPI,IKP]{Yu.~Valdau\corauthref{cor1}},
\ead{y.valdau@fz-juelich.de}
\corauth[cor1]{Corresponding author.}
\author[PNPI]{V.~Koptev},
\author[PNPI]{S.~Barsov},
\author[IKP]{M.~B\"uscher},
\author[DUBNA]{S.~Dymov},
\author[IKP]{M.~Hartmann},
\author[Tbilisi,Erlangen]{A.~Kacharava},
\author[IKP,DUBNA]{S.~Merzliakov},
\author[PNPI]{S.~Mikirtychyants},
\author[Erlangen]{A.~Mussgiller},
\author[PNPI,IKP]{M.~Nekipelov},
\author[IKP]{R.~Schleichert},
\author[IKP]{H.~Str\"oher}, and
\author[CW_College]{C.~Wilkin}

\address[PNPI]{High Energy Physics Department, Petersburg Nuclear
  Physics Institute, 188350 Gatchina, Russia}
\address[IKP]{Institut f\"ur Kernphysik, Forschungszentrum J\"ulich,
  52425 J\"ulich, Germany}
\address[DUBNA]{Laboratory of Nuclear Problems, Joint Institute for Nuclear Research,
  141980 Dubna, Russia}
\address[Tbilisi]{High Energy Physics Institute, Tbilisi State University, 0186 Tbilisi, Georgia}
\address[Erlangen]{Physikalisches Institut II, Universit\"at Erlangen--N\"urnberg, 91058 Erlangen, Germany}
\address[CW_College]{Physics and Astronomy Department, University College London, London, WC1E 6BT, UK}

\begin{abstract}
Inclusive $K^+$ production in proton-proton collisions has been
measured at a beam energy of 2.16\,GeV using the COSY-ANKE magnetic
spectrometer. The resulting spectrum, as well as those corresponding
to $K^+p$ and $K^+\pi^+$ correlated pairs, can all be well described
using consistent values of the total cross sections for the $pp\to
K^+p\Lambda$, $pp\to K^+p\Sigma^0$, and $pp\to K^+n\Sigma^+$
reactions. While the resulting values for $\Lambda$ and $\Sigma^0$
production are in good agreement with world data, our value for the
total $\Sigma^+$ production cross section,
$\sigma(pp\to{}K^+n\Sigma^+) =
(2.5\pm0.6_{\textrm{stat}}\pm0.4_{\textrm{syst}})\,\mu\textrm{b}$ at
an excess energy of $\varepsilon =129\,$MeV, could only be
reconciled with other recently published data if there were a highly
unusual near--threshold behaviour.
\end{abstract}

\begin{keyword}
Kaon production \sep Sigma production \sep Threshold effects

\PACS 13.75.-n   
\sep 14.20.Jn    
\sep 14.40.Aq    
\sep 25.40.Ve    

\end{keyword}
\end{frontmatter}

The production of light hyperons in proton--proton collisions in the
close--to--threshold region has been extensively studied at
different experimental facilities. The energy dependence of the
total cross sections for $pp{\to}K^+p\Lambda$ and $pp{\to}
K^+p\Sigma^0$ has been well measured and both follow phase--space,
though modified in the former case by the $p \Lambda$ final--state
interaction (FSI)~\cite{ls_tot_par}. On the other hand, little
information is available on the $pp{\to}K^+n\Sigma^+$ reaction. The
COSY-11 collaboration has recently published surprisingly high
values for the total cross sections in this channel at excess
energies of $\varepsilon = 13$\,MeV and 60\,MeV~\cite{cosy11}.
According to these measurements, the ratios of the total cross sections
$R(\Sigma^+/\Sigma^0)= \sigma(pp{\to}K^+n\Sigma^+)/
\sigma(pp{\to}K^+p\Sigma^0)$ at these two energies are $230{\pm}70$
and $90{\pm}40$, respectively~\cite{Tomasz}. These experimental
results are in striking contrast to published theoretical
estimates~\cite{Tsushima}. However, it has recently been suggested
that the inclusion in the production model of the previously ignored
$\Delta^{++*}(1620)1/2^{-}$ isobar, together with a strong
$n\Sigma^+$ FSI, would allow one to achieve much better (factor
2--4) agreement with the COSY-11 data~\cite{xie}.

A model--independent estimate for $R(\Sigma^+/\Sigma^0)$ might be
obtained from the isospin relation linking the different $\Sigma$
production channels, the amplitudes for which satisfy:
\begin{equation}
\label{amps}%
f(pp \to K^+ n \Sigma^+)+ f(pp \to K^0 p \Sigma^+) +
\sqrt{2}\,f(pp \to K^+ p \Sigma^0) = 0\,.
\end{equation}
This leads to a triangle inequality between the total cross
sections~\cite{Louttit}:
\begin{eqnarray}
\nonumber \left[\sqrt{\sigma(pp \to K^0 p
\Sigma^+)}-\sqrt{2\sigma(pp \to K^+ p \Sigma^0)}\,\right]^2\leq
\sigma(pp \to K^+ n \Sigma^+)\\ \leq \left[\sqrt{\sigma(pp \to K^0
p \Sigma^+)}+\sqrt{2\sigma(pp \to K^+ p \Sigma^0)}\,\right]^2\,.
\label{triangle}%
\end{eqnarray}

At $\varepsilon\approx129\,$MeV (the excess energy corresponding to
a proton beam energy of 2.16~GeV),
$\sigma(pp{\to}K^0p\Sigma^+)$~\cite{TOFK0} is nearly equal to
$\sigma(pp{\to}K^+p\Sigma^0)$~\cite{ls_tot_par} so that the
inequality of Eq.~(\ref{triangle}) predicts that
$R(\Sigma^+/\Sigma^0)<6$ at this excess energy. The COSY-11 results
exceed this limit by more than an order of magnitude, though they
were obtained closer to threshold, where no other $K^0p\Sigma^+$
data have been published\footnote{There are, however, data taken
with the COSY-TOF detector and presented in PhD
theses~\cite{TOF-theses}.}.

The authors of Ref.~\cite{sp_new} analysed published momentum
spectra from inclusive $K^+$ production in $pp$ collisions at
different angles and beam energies, with the aim of extracting the
contribution from the $K^+n\Sigma^+$ channel. For $K^+$ missing--masses
 below the $N\Lambda\pi$ threshold, only contributions from
the $K^+p\Lambda$, $K^+p\Sigma^0$, and $K^+n\Sigma^+$ channels are
relevant. It was assumed that production in the first two channels
could be described by three--body phase--space, with possible
modifications coming from the FSI. By subtracting these known
contributions from the inclusive spectra, an estimate of the
$pp{\to}K^+n\Sigma^+$ cross section was deduced. The inclusive data
available were restricted to relatively high excess energies,
$\varepsilon > 170$\,MeV, and had therefore no direct bearing on the
COSY-11 results. However, the authors did conclude that there was no
visible evidence for any strong $N \Sigma$ FSI.

Since one cannot \textit{a priori} exclude an anomalous threshold
behaviour associated with the isospin $I{=}\frac{1}{2}$ $K^+n$ (and
$K^0p$) system, as suggested in Ref.~\cite{xie}, further
experimental studies of the $pp{\to}K^+n\Sigma^+$ reaction are
necessary to clarify the situation. We here present the analysis of
new experimental data taken at a proton beam energy $T_p=2.157$\,GeV.

\begin{figure}[htb]
\centerline{\epsfxsize=2.3in\rotatebox{-90}{\epsfbox{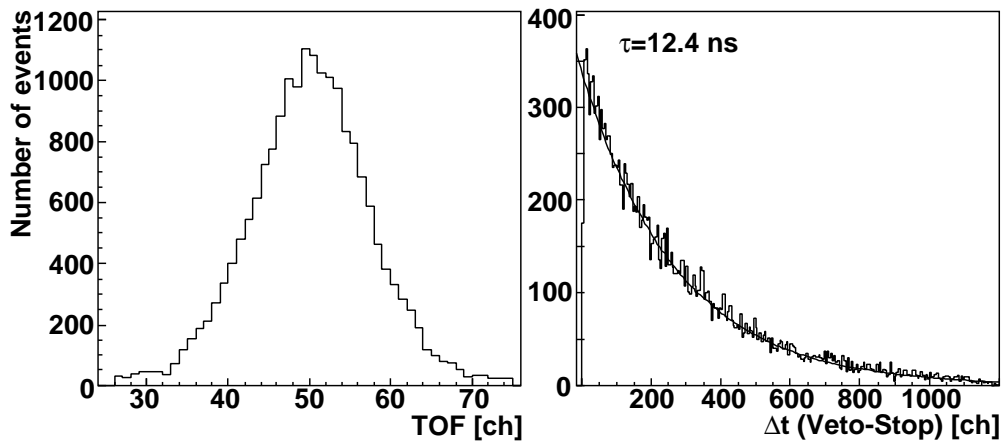}}}
\caption{\label{TOF}The time of flight (TOF) between start and stop
counters for inclusive $K^+$ production in $pp$ collisions at
$2.16\,$GeV (left panel). Time difference ($\Delta t$) between the
detection of the $K^+$ meson in the stop counter and a decay $\pi^+$
or $\mu^+$ in the corresponding veto counter of the same telescope
(right panel). The solid line, which corresponds to the 12.4\,ns
lifetime of the $K^+$, reproduces well the data. }
\end{figure}

The experiment was carried out using the magnetic spectrometer
ANKE~\cite{anke} at the COoler--SYnchrotron
COSY-J\"ulich~\cite{cosy}, with an internal cluster--jet target
which had an average density of
$\sim2{\times}10^{14}\,\textrm{cm}^{-2}$~\cite{clustertarget}. Only
two of the ANKE detection systems were needed for the analysis of
these data. The positive side system (PD), used for the $K^+$ and
$\pi^+$ detection, consists of 23 thin start counters, placed close
to the vacuum chamber window, two multiwire proportional chambers
(MWPCs), and 21 stop counters for time--of--flight (TOF)
measurements. The experimental efficiency of particle identification
was 98\% using time of flight and 90\% on average for MWPCs and was
known with an accuracy of $\sim1\%$. The first 15 stop counters are
part of range telescopes used for the identification of the $K^+$ mesons.
Each of these telescopes consists of a stop counter, energy--loss
counter, two passive degraders and a veto counter. The thickness of
the passive degraders in each telescope is chosen such that the
$K^+$ deposits the maximum energy in the energy--loss counter and
stops either at the edge of the counter or in the second passive
degrader. Delayed signals for the kaon decay products are then
registered by the so--called veto counter. This method (see
Fig.~\ref{TOF}) allows one to identify the $K^+$-mesons by
suppressing a background that is up to $10^6$ times higher. Such
data by themselves are sufficient for the determination of the
inclusive kaon spectrum. The efficiency of the kaon identification
by this method, which varies between 10--30\% depending on the
telescope number, is known with an accuracy of 10--15\%. Details of
the particle identification analysis using the delayed--veto
technique are to be found in Ref.~\cite{k_nim}.

The ANKE forward detector system (FD)~\cite{dymov} was used for both
the $K^+ p$ correlation measurements and luminosity determination.
The FD consists of two layers of plastic scintillator and a set of
three multiwire proportional chambers placed downstream of the
magnet. The efficiency of track reconstruction using FD MWPCs, 
which was about $85\%$, was known with an accuracy of approximately 
$1\%$. The luminosity was determined by selecting proton--proton 
elastic scattering events in the angular range 
$6.8^\circ < \theta_{\rm lab} < 8.8^\circ$
 on the basis of a dedicated pre--scaled trigger. This
is described in some detail in Ref.~\cite{Barsov}, where the same
data set was used for the investigation of $\omega$--meson
production. The overall systematic uncertainty in the absolute
normalisation was estimated to be of the order of
$6\%$~\cite{Barsov}. It is estimated that the amount of background
in the $K^+ p$ and $K^+ \pi^+$ correlation spectra is less then 2\%.
For the acceptance calculations, a model of the ANKE system has been
implemented within the GEANT4 simulation package~\cite{geant4}. This
contributes an overall uncertainty of about $5\%$.

Information on $\Sigma^+$ production was obtained from three
simultaneously measured observables, \textit{viz.}\ the $K^+p$,
$K^+\pi^+$ correlation spectra and the $K^+$ inclusive
double--differential cross section, which we first briefly outline.
The measured missing--mass spectrum of the detected $K^+p$ pairs
allows one to fix the strength of the different $K^+$ production
channels at this energy. Since the decay $\Sigma^+{\to}p \pi^0$ is
also possible (branching ratio BR 51.6\%), this spectrum also
contains some information on the $\Sigma^+$ production total cross
section, $\sigma(\Sigma^+)$.

The $pp{\to}K^+n\Sigma^+$ reaction can be cleanly identified either by
using $K^+n$ correlations, as at COSY-11~\cite{cosy11}, or by
detecting $K^+ \pi^+$ pairs coming from the decay
$\Sigma^+{\to}\pi^+n$ (BR 48.3\%). Although the
$pp{\to}K^+n\Lambda\pi^+$ reaction is another potential source of
$K^+\pi^+$ correlations, even at the much higher energy of 2.85\,GeV
its production is only about 4\% of that of
$\Sigma^+$~\cite{Louttit}. The contribution of this channel to the
final distributions is therefore estimated to be less than 2\%.

The inclusive $K^+$ double--differential cross section depends upon
all possible production channels, though the contribution from the
$pp{\to}K^+n\Sigma^+$ reaction at 2.16~GeV represents only a small
fraction of the total. Therefore, within our systematic errors, only
an upper limit for $\sigma(\Sigma^+)$ can be extracted from the
inclusive data at this energy. Nevertheless, this spectrum does
provide a valuable check on the consistency of the whole analysis by
using simulations where the individual weights of the channels are
fixed by the total cross sections extracted from the correlation
data.

\begin{figure}[htb]
\centerline{\epsfxsize=3.5in\rotatebox{-90}{\epsfbox{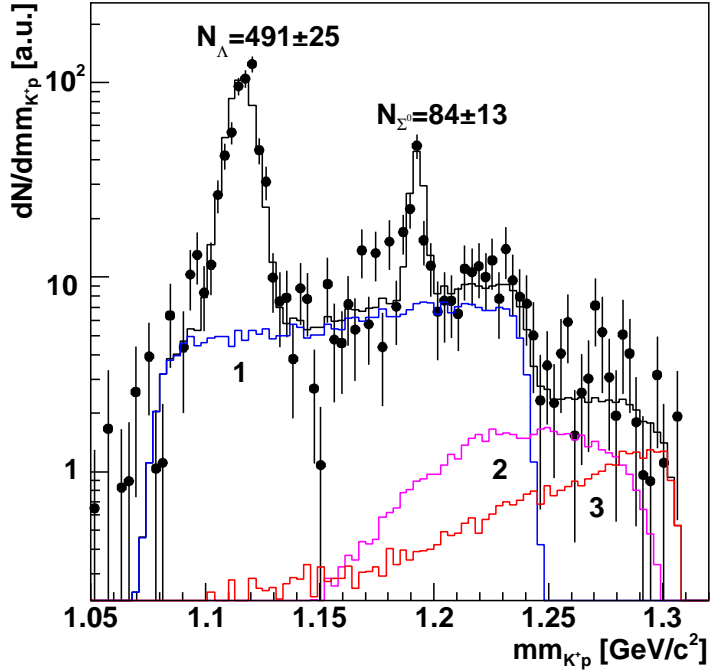}}}
\caption{\label{Kp-corr} Missing--mass distribution of $K^+p$ pairs ($mm_{K^+p}$) 
produced in $pp$ collisions at 2.16\,GeV. Experimental data are
shown by circles (resolution $\sim3$~MeV/$c^2$). The two peaks correspond
 to direct protons from the $pp \to K^+p \Lambda/\Sigma^0$ reactions. 
The continuum contributions of secondary protons arising from the 
$pp\to K^+p\,( \Lambda \to \pi^-p)$ (histogram 1), 
$pp \to K^+p(\Sigma^0\to \gamma \Lambda \to \gamma\pi^-p)$ 
(histogram 2), and $pp\to K^+n (\Sigma^+\to \pi^0 p)$ (histogram 3)
 have been obtained in Monte Carlo
simulations. The sum of all contributions, including the two direct
peaks, is shown by the solid histogram.}
\end{figure}

The $K^+p$ missing--mass spectrum presented in Fig.~\ref{Kp-corr}
shows two prominent peaks corresponding to $\Lambda$ and $\Sigma^0$
production. In addition there is a continuum resulting from the
detection of protons from the $\Lambda\to\pi^- p$ (BR 63.9\%) and
$\Sigma^0\to\gamma \Lambda \to \gamma p \pi^-$ (BR 100\%) decay, as
well as a contribution from the $\Sigma^+ \to p \pi^0$ decay. This
continuum is described well by our simulations.

Following the authors of Ref.~\cite{lambda285}, a simple model has
been developed for the $pp{\to}K^+p\Lambda$ reaction. We here assume
that %
(i) the $N^*(1650)$-resonance is the dominant contribution for
$\Lambda$ production, %
(ii) the $p \Lambda $ FSI~\cite{ls_tot_par} has a significant effect
on the experimental observables, %
(iii) use the angular distribution of the vertex-proton, as measured
with the COSY-TOF detector~\cite{ls0295}. %
A simple phase--space model has been used for the
$pp{\to}K^+p\Sigma^0$ and $pp{\to}K^+n\Sigma^+$ reactions since
there is no evidence for significant $p \Sigma^0$ or $n \Sigma^+$
FSI effects~\cite{ls_tot_par, sp_new}, and this is confirmed by our
data.

The number of events extracted from the measured missing--mass
spectrum of Fig.~\ref{Kp-corr}, together with our values for the
total acceptances, luminosity, and efficiencies, yields total cross
sections of
$\sigma(\Lambda)=(23.2\pm3.7_{\textrm{stat}}\pm5.8_{\textrm{syst}})\,\mu$b
and
$\sigma(\Sigma^0)=(2.6\pm0.6_{\textrm{stat}}\pm0.4_{\textrm{syst}})\,\mu$b
for $\Lambda$ and $\Sigma^0$ production, respectively. For the total cross
sections calculations only the direct $K^+p$ events are used, as
their amount is precisely known (peaks in Fig.~\ref{Kp-corr}). These values
are in agreement with the parameterisation of the world data 
(see Ref.~\cite{ls_tot_par}).

The high--mass part of the missing--mass spectrum in the
Fig.~\ref{Kp-corr} is sensitive to $R(\Sigma^+/\Sigma^0)$. A good 
description of the spectrum can be obtained if $R(\Sigma^+/\Sigma^0)
\approx 1.5$. However our statistics do not permit us to draw
meaningful conclusions on the associated error.

\begin{figure}[htb]
\centerline{\epsfxsize=2.3in\rotatebox{-90}{\epsfbox{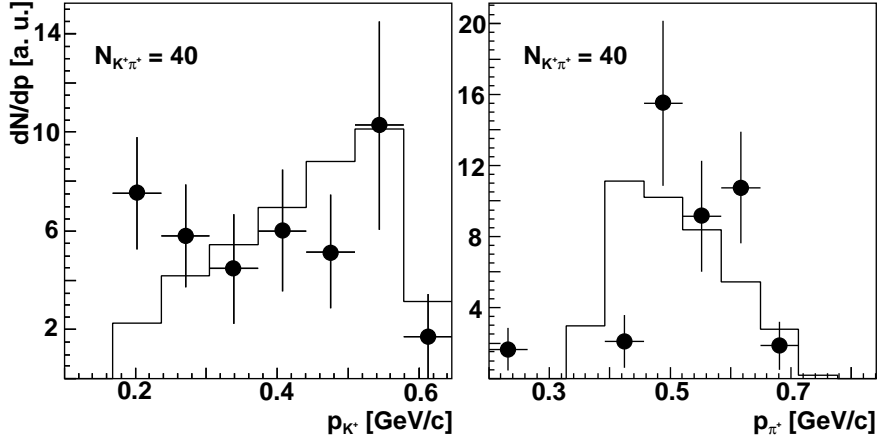}}}
\caption{\label{Kpi} Momentum distributions of $K^+$ and $\pi^+$
from the $pp\to K^+\pi^+X$ reaction at 2.16\,GeV. The solid
histograms correspond to simulations of the $pp\to K^+ n (\Sigma^+\to \pi^+n)$
reaction in the phase--space model. }
\end{figure}

The momentum distributions of the detected $K^+ \pi^+$ pairs
are presented in Fig.~\ref{Kpi}. Simulations carried out within the
framework of the phase--space model show reasonable agreement with
the experimental data. From the number of detected events the total
cross section is determined to be:
\[
\sigma_{\rm tot}(\Sigma^+)
=\left(2.5\pm0.6_{\textrm{stat}}\pm0.4_{\textrm{syst}}\right)
\,\mu\textrm{b},
\]
where both the statistical and systematic uncertainties are
indicated.

The ratio of the $\Sigma^+$ and $\Sigma^0$ count rates,
$N_{\Sigma^+}/N_{\Sigma^0}$, is practically independent of the
conditions of the experiment (luminosity, telescope efficiencies
\textit{etc.}). It can therefore be used to cross check the
experimental value of $\sigma(\Sigma^+)$ extracted from the analysis
of $K^+\pi^+$ correlations. The $\sigma(\Sigma^+)/\sigma(\Sigma^0)$
ratio depends on the acceptances ($A$) and number of detected events 
($N_{K^+\pi^+}$ for the $\Sigma^+$, and $N_{K^+p}$ from direct proton
for $\Sigma^0$):
\begin{equation}
\frac{\sigma(\Sigma^+)}{\sigma(\Sigma^0)}=
\frac{N_{K^+\pi^+(\Sigma^+)}}{N_{K^+p(\Sigma^0)}} \times
\frac{\textrm{A}_{K^+p(\Sigma^0)}}{\textrm{A}_{K^+\pi^+(\Sigma^+)}}
\times\frac{1}{\textrm{BR}_{\Sigma^+\to\pi^+n}}
\end{equation}

Using the numbers of events extracted from the experimental spectra 
together with our estimates of the total acceptances, we obtain the
following ratio of the $\Sigma^+/\Sigma^0$ total cross sections:
\begin{equation}
\label{Sigratio} \frac{\sigma(\Sigma^+)}{\sigma(\Sigma^0)}
=\frac{(40\pm7)}{(84\pm13)}\times\frac{4.5\times
10^{-4}}{5.1\times10^{-4}}\times\frac{1}{0.48}=0.9\pm0.2\:.
\end{equation}

\begin{figure}[htb]
\centerline{\epsfxsize=2.8in\rotatebox{-90}{\epsfbox{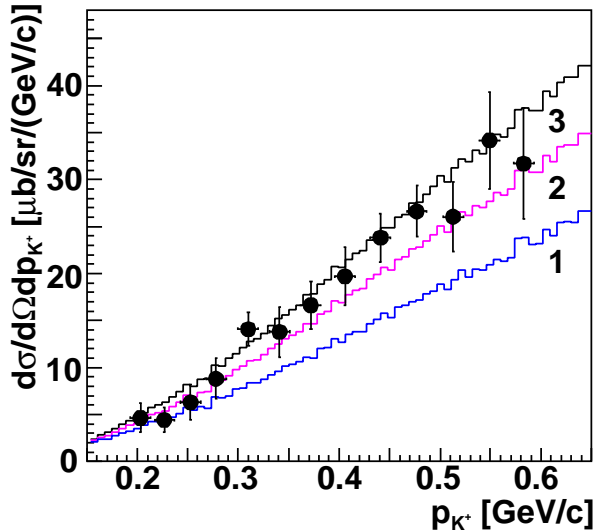}}}
\caption{\label{inclusive}Inclusive $K^+$ momentum spectrum for
$\theta_K^{\rm lab}<4^{\circ}$ resulting from $pp$ collisions at
2.16\,GeV. The simulation of $pp{\to}K^+p\Lambda$ with
$\sigma(\Lambda)=23.2\,\mu\textrm{b}$ is shown by the histogram 1.
The addition of the contribution from the $pp{\to}K^+p\Sigma^0$
reaction using a total cross section of $\sigma(\Sigma^0)=2.6\,\mu\textrm{b}$
 leads to the histogram 2. The total, corresponding to the further
inclusion of the $pp{\to}K^+n\Sigma^+$ reaction channel with
$\sigma(\Sigma^+) = 2.5\,\mu\textrm{b}$, is shown by the histogram 3.
}
\end{figure}

Since the ratio is consistent with unity, the $\Sigma^+$ total cross section
 derived from Eq.~(\ref{Sigratio}) agrees with the value obtained directly
 from the $K^+\pi^+$ data, as well as that estimated from the
 $K^+p$ missing--mass spectrum. It is also
reassuring that our simulation of the inclusive $K^+$ spectrum shown
in Fig.~\ref{inclusive} reproduces the experimental data  so well.
This means that the relation between the inclusive and correlation
data seems to be well understood.

\begin{figure}[htb]
\centerline{\epsfxsize=2.8in\rotatebox{-90}{\epsfbox{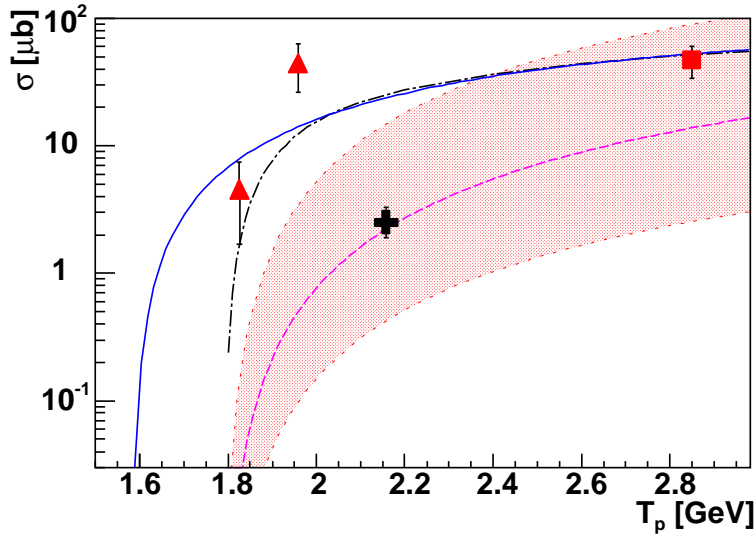}}}
\caption{\label{sigtot}%
Total cross sections as a function of beam energy. Our value for
$\sigma(pp{\to}K^+n\Sigma^+)$ at 2.16\,GeV is shown by a cross.
Total cross sections for $pp{\to}K^+n\Sigma^+$ reaction measured by
COSY-11~\cite{cosy11} and at higher energy in a bubble--chamber
experiment~\cite{Louttit} are represented by triangles and an
 square, respectively. The lines show the normalised three--body
phase space dependence for the $pp{\to}K^+p\Lambda$ (solid line) and
$pp{\to}K^+p\Sigma^0$ (dashed line) with FSI effects in the
$\Lambda$ case, as described in Ref.~\cite{ls_tot_par}. Both
reproduce well the available experimental data. The estimate for the
$pp{\to}K^+n\Sigma^+$ total cross section from Ref.~\cite{xie} is
shown by the chain curve. The region restricted by the triangle
inequality of Eq.~(\ref{triangle}) is shown by the hatched area.}
\end{figure}

Our value of the $\Sigma^+$ production cross section falls well
within the boundaries fixed by isospin invariance that are shown in
Fig.~\ref{sigtot}. It is also in agreement with experimental data
collected with the COSY-TOF detector at the same
energy~\cite{SPTOF-theses}. Compared to this the two COSY-11 points,
which were taken even closer to the threshold, look extremely high.

In summary, we have presented new measurements of the
$pp{\to}K^+n\Sigma^+$ total cross section at 2.16~\,GeV that do not
dependent upon the detection of the final neutron. From the analysis
of the $K^+ p$ and $K^+ \pi^+$ correlated pairs, total cross
sections for the production of $\Lambda$, $\Sigma^0$ and $\Sigma^+$
have been extracted. The values of $\sigma(\Lambda)$ and
$\sigma(\Sigma^0)$ are in reasonable agreement with the trends of
the experimental data defined at other energies. Our value of
$\sigma(\Sigma^+)$ at $\varepsilon =129$\,MeV satisfies well the
triangle inequality of Eq.~(\ref{triangle}). Furthermore, the
inclusive double--differential cross section is well described using
the values of the total cross sections for the individual $K^+$
production channels determined in this work from the correlation
studies. This shows an overall consistency of the methodology.

Our data show that at $\varepsilon \approx 128$\,MeV the $\Sigma^+$
and $\Sigma^0$ production rates are rather similar and the
expectation would be that this would continue as the threshold is
approached. However, the value of the total $\Sigma^+$ cross section
reported by the COSY-11 collaboration at $\varepsilon =60$\,MeV is
over an order of magnitude larger than ours. Taken at face value,
the two measurements would imply a very large threshold anomaly.
Even if this seems to be very unlikely, it can and must be checked,
and this is possible with our method~\cite{pr171}.

This work has been partially supported by BMBF, DFG(436 RUS
113/768), Russian Academy of Science, and COSY FFE. We thank the
COSY machine crew for their support during the experiment.
 J.-J.~Xie and B.-S.~Zou kindly provided us with the
numerical values of the calculations in Ref.~\cite{xie}.
Useful discussions with A.~Sibirtsev and other members of the
ANKE Collaboration are gratefully acknowledged.
%
%

\end{document}